\documentclass[aps, reprint, showpacs, superscriptaddress, pre, citeautoscript]{revtex4-1}

\usepackage{amsmath}
\usepackage{amssymb}
\usepackage{graphicx}
\usepackage{caption}
\usepackage{subcaption}
\usepackage{hyperref}
\usepackage{xcolor}
\usepackage{gensymb}

\usepackage{verbatim}

\newcommand{\etal}{{\it et al.}}
\newcommand{\authornote}[1]{{\color{red} #1}}

\usepackage{soul}

\begin{document}
\title{Inferring low-dimensional microstructure representations using
  convolutional neural networks}

\author{Nicholas Lubbers}
\email{nlubbers@bu.edu}
\affiliation{Department of Physics, Boston University, Boston, MA}
\affiliation{Theoretical Division and CNLS, Los Alamos National
  Laboratory, Los Alamos, NM 87545}

\author{Turab Lookman}
\affiliation{Theoretical Division and CNLS, Los Alamos National
  Laboratory, Los Alamos, NM 87545}

\author{Kipton Barros}
\affiliation{Theoretical Division and CNLS, Los Alamos National
  Laboratory, Los Alamos, NM 87545}

\date{\today}

\begin{abstract}
We apply recent advances in machine learning and computer vision 
to a central problem in materials informatics: The statistical
representation of microstructural images. We use activations in a
pre-trained convolutional neural network to provide a
high-dimensional characterization of a set of synthetic
microstructural images.  Next, we use manifold learning to obtain a
low-dimensional embedding of this statistical characterization. We
show that the low-dimensional embedding extracts the
parameters used to generate the images. According to a
variety of metrics, the convolutional neural network method yields
dramatically better embeddings than the analogous method derived from two-point correlations alone.
\end{abstract}

\maketitle

\section{Introduction}

A central problem in materials design is the analysis,
characterization, and control of materials
microstructure. Microstructure is generated by non-equilibrium
processes during the formation of the material and plays a large role
in the bulk material's
properties~\cite{Kumar06,Ostoja-Starzewski07,Wang08,Fullwood10,Torquato10}.
In recent years, machine learning and informatics based approaches to materials design have
generated much
interest~\cite{Rajan13,Kalidindi15,Lookman16a,Voyles17,Xue16}. Effective
statistical representation of microstructure has emerged as an outstanding
challenge~\cite{Kalidindi11,Liu13,Niezgoda13}.

Standard approaches begin with an $n$-point expansion, and typically truncate at the pair
correlation level~\cite{Jiao07,Fullwood08,Jiao09,Chen14}. Pair correlations can capture information such as the
scale of domains in a system, but miss higher order complexities such as the
detailed shape of domains or relative orientation of nearby domains~\cite{Jiao10a,Jiao10,Gommes12,Gommes12a}.
Three-point correlations (and successors) quickly become
computationally infeasible, as the number of $n$-point correlations
scales exponentially with $n$. Furthermore, they are not tailored to capture the statistical
information of interest. Much current work involves
deploying a set of modified two-point correlations to better capture
certain microstructural
features~\cite{Jiao09,Niezgoda08,Zachary11,Gerke14,Xu14}. 

Independently, researchers in machine learning for computer vision
have been developing powerful techniques to analyze image
content~\cite{Lecun98,Krizhevsky12,Simonyan14,Szegedy15,He16}. Deep
Convolutional Neural Networks (CNNs) have emerged as a particularly
powerful tool for image analysis~\cite{LeCun15}. Of
particular interest to materials microstructure analysis is literature
regarding \emph{texture} reconstruction and modeling~\cite{Hao12,Kivinen12,Luo13,Gao14,Theis15}; in this context a
texture is an image with roughly translation invariant statistics.
Indeed, Gatys et al. have recently demonstrated that correlations between CNN activations capture the statistics of textures exceptionally well~\cite{Gatys15a,Gatys16}. 

\begin{figure*}
  \centering \includegraphics[width=1.\textwidth]{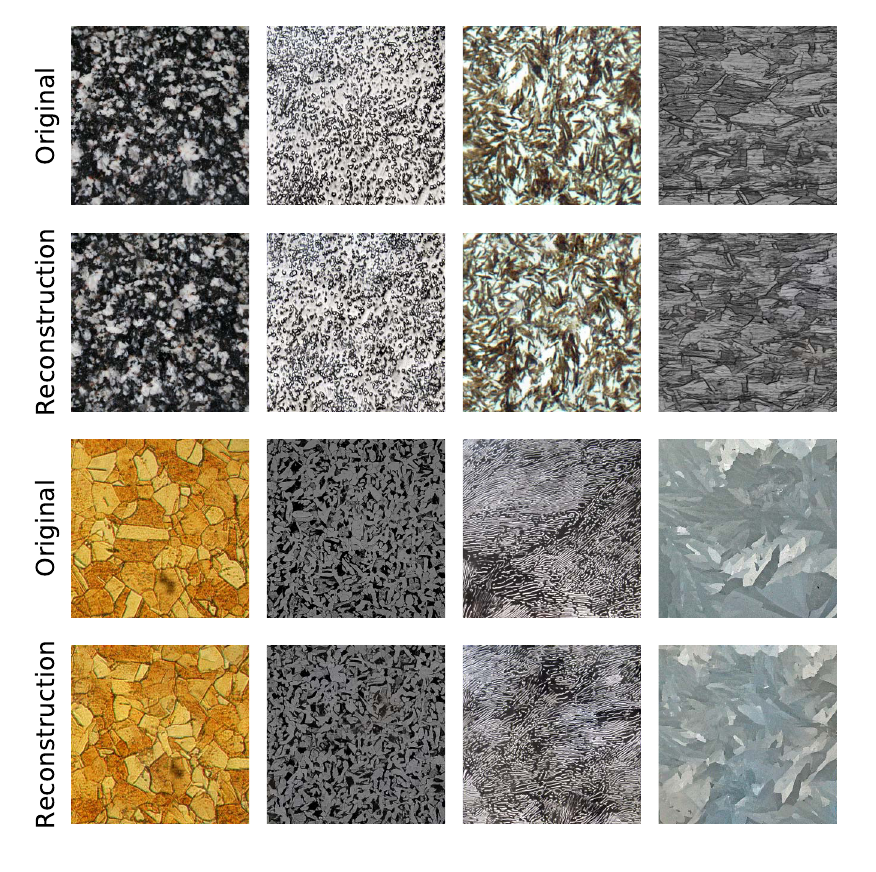}
  \caption{(Color online) Texture synthesis of materials microstructures using
    the CNN algorithm from~\cite{Gatys15a}. The CNN synthesizes each ``Reconstruction''
    image from a single ``Original'' image. (Image attributions listed in Appendix~\ref{sec:imagecredits}.)}
  \label{fig:texturereconstructions}
\end{figure*}

Here, we apply the Gatys et al. CNN {\em texture vector}
representation of image statistics to the problem of characterizing
materials micrographs. The texture vector is a
statistical representation of an image derived from the processing of a pre-trained
CNN; we use the
Visual Geometry Group's VGG-19 network~\cite{Simonyan14}, which has
been trained to classify 1.2 million natural
images~\cite{Russakovsky15}. We
  demonstrate that the texture vectors generated using the VGG-19
  network can capture complex statistical correlations visible in
  microstructure images in Fig.~\ref{fig:texturereconstructions}.
Specifically, we use the CNN texture vector to characterize an original microstructural image, and then generate a new, random image constrained to the same statistics.
It is remarkable that using only a single original image, the algorithm generates texture images nearly indistinguishable to the eye.
In the case of materials micrographs, where data can be
  expensive to collect, the ability for a method perform well on small
  datasets is crucial.  Our approach can be considered one of \emph{transfer
    learning}, \emph{i.e.}, the application of a model trained on one
  problem to achieve results on a different, but related problem. The
fidelity in Fig.~\ref{fig:texturereconstructions} motivates us to
pursue the CNN texture vector as a tool for mapping relationships
between materials processing, microstructure, and properties. 

As a step towards microstructure analysis, we
demonstrate and quantify the ability of the CNN texture vector to
extract hidden information from a collection of synthetic
texture images. Images in our datasets are generated under
different ``processing'' conditions, \emph{i.e.} a few generating
parameters. The goal is to extract a compact statistical
representation of the dataset that captures the relevant statistics
associated with the hidden generating parameters---that is, to back out
the processing parameters of micrographs directly from the
images alone. Our use of synthetic data, with a tightly controlled ground truth,
provides us quantitative measures of error in the unsupervised learning process.
  
Despite being very
high-dimensional, the CNN texture vector offers a good notion of
abstract distance between texture images: The closer the generating parameters, the smaller the distance between texture vectors.
We use \emph{manifold learning} to embed each image as a point in a low-dimensional space such that, ideally, the
embedded distances match the texture vector distances. The structure of the embedded points then reveals information about the set of images. For our synthetic dataset, we show a simple relationship between the embedded coordinates and the generating parameters.
More broadly, dimensionality reduction techniques may
serve as the basis for characterization of materials properties
that are controlled by complex materials microstructures. A recent example is the work of Ref.~\onlinecite{DeCost17}, which uses a method similar to ours to map the space of ultrahigh carbon steel microstructures.

Our approach applies \emph{unsupervised} learning, a
pattern-discovery framework to seek new aspects of
microstructure without using labeled data. This approach is applicable
  to problems where the ground truth is unknown, \emph{e.g.}, the forensic analysis of microstructures. Several recent applications of machine learning to microstructure have used \emph{supervised} learning
algorithms~\cite{Sundararaghavan05,DeCost15,Kalinin15,Liu15,Bostanabad16,Chowdhury16,Orme16}
such as support vector machines and classification trees, which make inferences based on labeled data.
Like our work, Ref.~\onlinecite{Chowdhury16} uses image features
extracted from CNNs to aid microstructure analysis.
  
The remainder of the paper is organized as follows: Section~II gives
background on recent CNN architectures for image recognition, and specifics of the VGG network. Section~III details
our algorithms for statistical microstructure analysis and Sec.~IV
evaluates their accuracy on test datasets. Section~V provides
discussion and interpretation of our results, and we conclude in Sec. VI.

\section{Review of Convolutional Neural Networks}

CNNs have emerged in recent years as
state-of-the-art systems for computer vision
tasks~\cite{Lecun98,Krizhevsky12,Simonyan14,Szegedy15}.
They form a modern basis for image recognition and object detection
tasks, and in some cases now outperform humans~\cite{He15a,He16}.

The basic computational structure is that of a many-layered
  (\emph{i.e.}, {\em deep}) artificial neural network. For a brief overview, see Ref.~\onlinecite{LeCun15}; for a comprehensive text,
see Ref.~\onlinecite{Goodfellow16}.  There are a great variety of deep
neural network architectures; here we first focus on the core components. Each layer in the network contains many computational units ({\em neurons}).
Each neuron takes a set of inputs from
the previous layer and computes a single output (the {\em activation}) to be used as an
input in the next layer. Each neuron's activation is constructed as follows: First,
the set of inputs is linearly combined into a scalar using
a set of \emph{weights} and shifted using a \emph{bias}. To this sum, the neuron applies a simple
nonlinear map, the \emph{activation function}, to
generate its activation.

In the learning phase, the network is \emph{trained} by iteratively tuning the
weights and biases so that the network
better performs a task. Performance of the network is quantified by a scalar
\emph{objective function}.
Commonly, a network is trained by \emph{supervised learning}, in which the
network learns a mapping from inputs to outputs using a database of training
examples. In this case, the objective function is a measure of error
in the network output summed over all examples of the training
set. The objective function is often differentiable and optimized via \emph{stochastic
  gradient descent}.

A CNN is a specific type of artificial neural
  network which is useful for processing data on a spatial and/or
  temporal grid. The \emph{convolutional layers} in CNNs impose strong
  restrictions on the structure of weights: Each layer consists of a
  bank of trainable filters (sometimes called \emph{kernels}) that are
  convolved with activations from the previous layer. The convolution
  outputs are called {\em activation maps}. This technique of
  constraining and reusing weights is called {\em weight tying}. Note
  that the convolutional structure preserves spatial locality: The
  activation maps at each convolutional layer are interpretable as
  images. As in a plain artificial neural network, each pixel in the
  output image is passed through a nonlinear activation function. CNNs
  also commonly include {\em pooling layers} that effectively
  coarse-grain the image plane. These layers operate by taking a
  statistic over a small region of the image plane, such as the
  maximum of a feature's activations in a $2\times2$ pixel
  region. Importantly, the convolutional and pooling layers process
  the input image in a (nearly) translation equivariant way. This
  directly encoded translational symmetry is designed to match that of
  natural images, which as a distribution exhibit repeated patterns
  centered at a variety of locations. By alternating between sets
  of convolution and pooling layers, CNNs are able to develop
  sensitivity to very complex correlations over large length scales,
which underlies their strong performance on image recognition tasks.

As in Gatys \etal~\cite{Gatys15a,Gatys16}, our work begins with a normalized
version of the Visual Geometry Group's VGG-19 network~\cite{Simonyan14} already trained to classify natural images.
The VGG-19 network placed first in
localization and second in classification in the ILSVRC 2014 ImageNet
Challenge~\cite{Russakovsky15}. The VGG network is known for its simple architecture and
competitive performance. The convolutional kernels each have
a $3\times3$ pixel spatial extent. The nonlinear
activation function applied after each convolution is a rectifier
({\em ReLU}), $f(x) = \max(0, x)$. The convolutional layers are applied in a series
of blocks, and between the blocks, pooling layers are applied (in the
original network, Max pooling, but here as in
Refs.~\onlinecite{Gatys15a} and~\onlinecite{Gatys16} we use Mean pooling). Blocks
one and two contain two convolutions each, and blocks three, four, and five
contain four convolutions each. The final stage of the network
adds three fully connected layers---these do not
directly encode spatial information, and so are not used for
translation invariant characterization of images. We used the
optimizing compiler Theano~\cite{Theano-Development-Team16} and the
neural network library Lasagne~\cite{Dieleman15} to implement the
CNN methods used in this paper.

\section{Methods}

\subsection{CNN texture vector representation of image statistics}

\label{sec:cnntextureprocedure}

\begin{figure*}
  \centering \includegraphics[width=1.\textwidth]{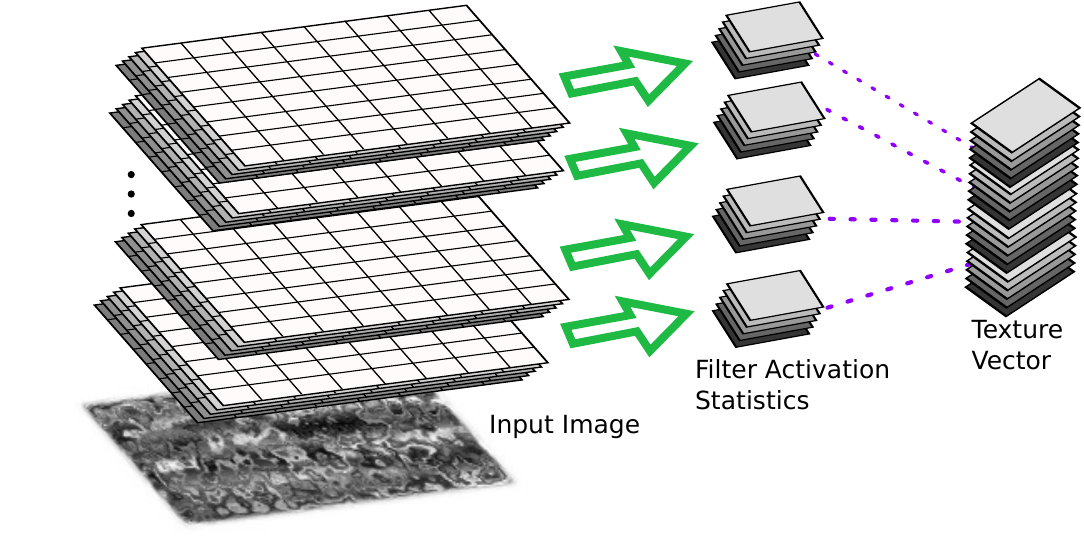}
  \caption{A schematic view of the creation of a texture vector
    representation of an image using a CNN. The input image is processed upward through
    multiple CNN layers. Activations at each layer produce a stack of filtered images. Correlations between filtered images, averaged over the image plane,
    are collected. These
    statistics are then concatenated (with optional weighting) to form
    a texture vector. The activations at higher layers capture texture information over larger spatial scales.}
\label{fig:texturescheme}
\end{figure*}

Gatys et al.~\cite{Gatys15a,Gatys16} have developed a robust algorithm for statistical analysis of texture images relevant to materials microstructure, which we demonstrated in Fig.~\ref{fig:texturereconstructions}.
Given an input texture image, the Gatys procedure extracts a texture vector from activations in the convolutional layers. This calculation is summarized in Fig.~\ref{fig:texturescheme}. Activations in the CNN are denoted by $F_{ij}^l$, where $l$ is the layer index, $i$ is the feature index, and $j$ is the pixel index. At each layer, the Gram matrix
captures correlations between feature $i$ and feature $k$,
\begin{equation}
G^l_{ik}  = \sum_j F_{ij}^lF_{kj}^l.
\end{equation}
The summation over pixel index $j$ encodes invariance to translations in the image plane (up to boundary effects).
Compared to the mean feature activations $\sum_j F_{ij}^l$ alone, the Gram matrix offers much richer statistical information~\cite{Gatys15a}. 
In the following we suppress feature indices $i$ and $k$ and view
  the Gram matrix, $G^l_{ik} \rightarrow G^l$, as a summary of
  activation statistics on layer $l$.
For the purposes of texture synthesis, Gatys et al. introduce a scalar,
positive-definite loss between two images $x_i$ and $x_j$ (let us note
explicitly that $i$ and $j$ now index images) with Gram matrices
$G^l(x_i)$ and $G^l(x_j)$:
\begin{equation}
\label{eqn:layerloss}
L^l(x_i,x_j) = \frac{1}{A_l} ||G^l(x_i)-G^l(x_j)||^2,
\end{equation}
with $||\cdot||$ the Euclidean norm.
$A_l = 4 N_l^2 M_l^2$ is a normalization factor for the loss on layer
$l$ with $N_l$ features and $M_l$ pixels.
The total loss is the weighted sum of layer-wise losses,
\begin{equation}
L(x_i,x_j) = \sum_l w_i L^l(x_i,x_j).
\end{equation}
In this work we use the VGG network~\cite{Simonyan14}, normalized as
in~\cite{Gatys15a,Gatys16}, and apply equal weight ($w_l=1$) to each
of the following layers: ``conv1\_1'', ``conv2\_1'', ``conv3\_1'',
``conv4\_1'', and ``conv5\_1''. 
  It is convenient to define rescaled Gram matrices, $\hat G^l =
  (w_l/A_l) G^l$. Their concatenation $\hat G = (\hat G^1, \hat G^2,
  \cdots)$ is the scaled texture vector. The layers we use have
  feature sizes of $[N_1, N_2, \hdots] = [64,128,256,512,512]$, resulting in a
  total texture vector length of $\sum_l N_l^2 \approx 5\cdot10^5$
  elements. From the texture vector of two images, we may form a
distance between images
\begin{equation}
\label{eqn:layerwisedist}
d_{ij}^2 = ||\hat G(x_i) - \hat G(x_j)||^2 = L(x_i,x_j).
\end{equation}
That is,  $\hat G$ as a function
  endows two images $x_i$ and $x_j$ with a Euclidean distance $d_{ij}$
  based on their texture representations within the CNN. We will show that this distance
  is a useful input to machine learning
algorithms, and in particular, manifold learning (see Sec.~\ref{sec:MDSintro}).

\subsection{Power spectrum statistics}

\label{sec:pstextureprocedure}

To benchmark the CNN texture vector representation, we compare it against the power
spectrum (PS) associated with two-point correlations in the image. This approach is commonly employed for statistical
characterization of microstructure. Our test dataset contains single-component
(grayscale) images $x_i$, each represented as a scalar field $\phi_i(\mathbf r)$. Assuming translation invariance, the
two-point correlation function of $\phi(\mathbf r)$ is
\begin{equation}
P_2(\Delta \mathbf r) = \int  \phi(\mathbf r') \phi(\mathbf r' +\Delta\mathbf r)
~ \mathrm{d}\mathbf r'.
\end{equation}
If ensemble averaged, the full set of $n$-point correlation
functions would capture all information about the statistical distribution of images.

The PS (also known as the structure factor) $S(\mathbf q)$ is the Fourier transform of $P_2$, which can be expressed as
\begin{equation}
S(\mathbf q) = \tilde{\phi}(\mathbf q)\tilde{\phi}(-\mathbf q) = |\tilde \phi(\mathbf q)|^2,
\end{equation}
where  $\tilde{\phi}(\mathbf q)$ is the Fourier transform of $\phi(\mathbf r)$.
For this analysis, we compute the PS after rescaling
$\phi(\mathbf r)$ to the range $[-1, 1]$.

To develop low-dimensional representations of microstructures, we
require a distance between microstructures.
Given two images $x_i$ and $x_j$ and their respective power spectra $S_i(\mathbf q)$ and $S_j(\mathbf q)$, we obtain a new distance $d_{ij}$ between the images,
\begin{equation}
\label{eqn:structuredist}
d_{ij}^2 = \int \left[ S_i(\mathbf q)- S_j(\mathbf q)\right]^2 \mathrm{d} \mathbf q,
\end{equation}
which should contrasted with the CNN distance in Eq.~\eqref{eqn:layerwisedist}.

\subsection{Manifold Learning with Multidimensional Scaling}
\label{sec:MDSintro}

To assess the quality of the texture vectors for characterizing
images we perform manifold learning. The goal of manifold learning is to find a low-dimensional
representation of the data that faithfully captures distance information  [here, from Eq.~\eqref{eqn:layerwisedist} or~\eqref{eqn:structuredist}]. Multidimensional
Scaling~\cite{Kruskal64,Borg05,Franklin08} (MDS) implements this principle
as follows: Given a dataset $\{x_i\}$, a distance function $d_{ij} = d(x_i, x_j)$ between datapoints, and an embedding dimension $D^\ast$, the
goal is to find embeddings $x_i \mapsto \hat x_i \in \mathbb{R}^{D^\ast}$ such that the embedded Euclidean distance $\hat d_{ij} = || \hat x_i - \hat x_j ||$ best matches $d_{ij}$, \emph{i.e.} minimizes a stress function $\sigma$. Here, we use Kruskal's stress~\cite{Kruskal64},
\begin{equation}
\label{eq:MDSstress}
\sigma = \sqrt{ \frac{\sum ( d_{ij} - \hat{d}_{ij} )^2}{  \sum    d_{ij}^2  } }
\end{equation}
We will consider embedding dimensions $D^\ast \lesssim 10$, which is very small compared to the dimension of the full CNN texture vector, roughly $2.5 \times 10^5$ (Sec.~\ref{sec:cnntextureprocedure}).

Note that MDS seeks $\hat d_{ij}$ that globally matches
distances $d_{ij}$, and thus captures information about \emph{all} image pairs. Other schemes, such as
Local Linear Embedding~\cite{Roweis00} or Isomap~\cite{Tenenbaum00},
instead work with a local sparsification of the distance matrix
$d_{ij}$. In this work, we select MDS because of its direct interpretibility and conceptual simplicity. MDS requires only one hyperparameter, the embedding dimension $D^\ast$.
We use the scikit-learn~\cite{Pedregosa11} implementation of MDS, which applies an iterative majorization algorithm~\cite{De-Leeuw77} to optimize the embedding stress $\sigma$, Eq.~\eqref{eq:MDSstress}.

\section{Tasks}

\subsection{Image generation process}

\begin{figure}
  \centering
    \includegraphics[width=0.5\textwidth]{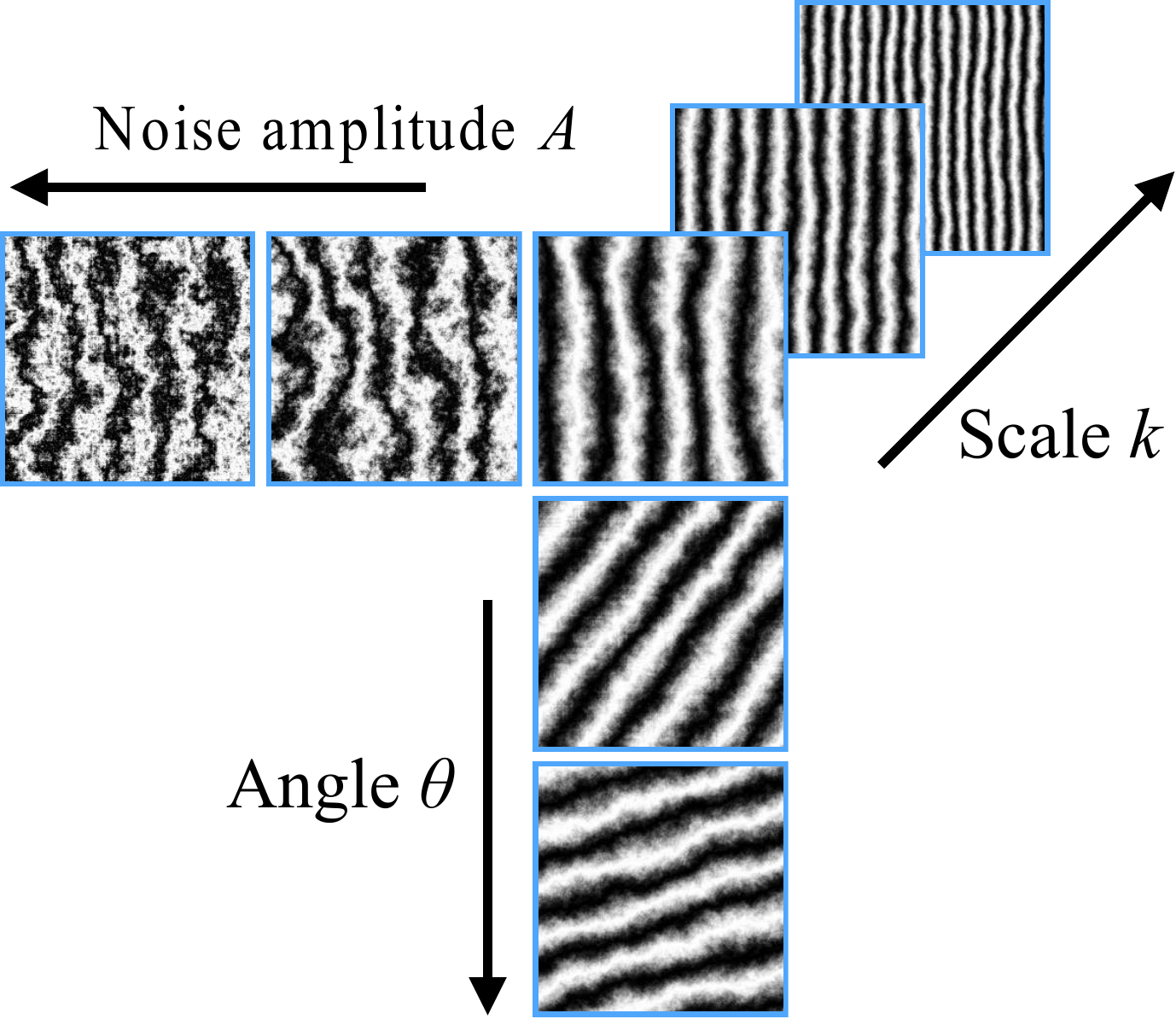}
  \caption{The space of synthetic textures, generated by tunable parameters $A$, $k$, and $\theta$.}
  \label{fig:3DManifoldScheme}
\end{figure}

We argue that although the space of materials microstructure is very rich, it will admit an effective low-dimensional representation.
For example, a description of the materials processing (\emph{e.g.}
composition, thermodynamic variables and their rates of change) should be more compact than a direct description of the resulting microstructure. A statistical analysis of microstructure is valuable in that it may lead to further dimensionality reduction; multiple different processing paths may lead to the same microstructure.
In this work, we study a database of synthetic 2-D microstructure images generated from a stochastic process with a few tunable generating variables.
We use Perlin noise~\cite{Perlin85} to generate marble-like stochastic images. The method procedurally
calculates a smooth multi-scale noise function $\mathbf h(\mathbf r)$
that generates distorted spatial points $\mathbf u_A = \mathbf r \, +\, A \, \mathbf h(\mathbf r)$ with noise amplitude $A$. Then each texture image $x_i$ is realized as a 2-D scalar field
\begin{equation}
\label{eq:textures}
\phi_i(\mathbf r) = \cos [2 \pi k_i\, \mathbf u_{A_i} \cdot  \mathbf {\mathbf{\hat n}}_{\theta_i}],
\end{equation}
where $\mathbf{\hat n}_\theta$ is a unit vector with angle $\theta$. That is, each image $x_i$ consists of
sinusoidal oscillation parameterized by angle ($\theta_i$), scale ($k_i$), and noise amplitude ($A_i$) parameters.
This three-dimensional parameter space is shown in Fig.~\ref{fig:3DManifoldScheme}.

\begin{figure*}
  \centering
    \includegraphics[width=0.8\textwidth]{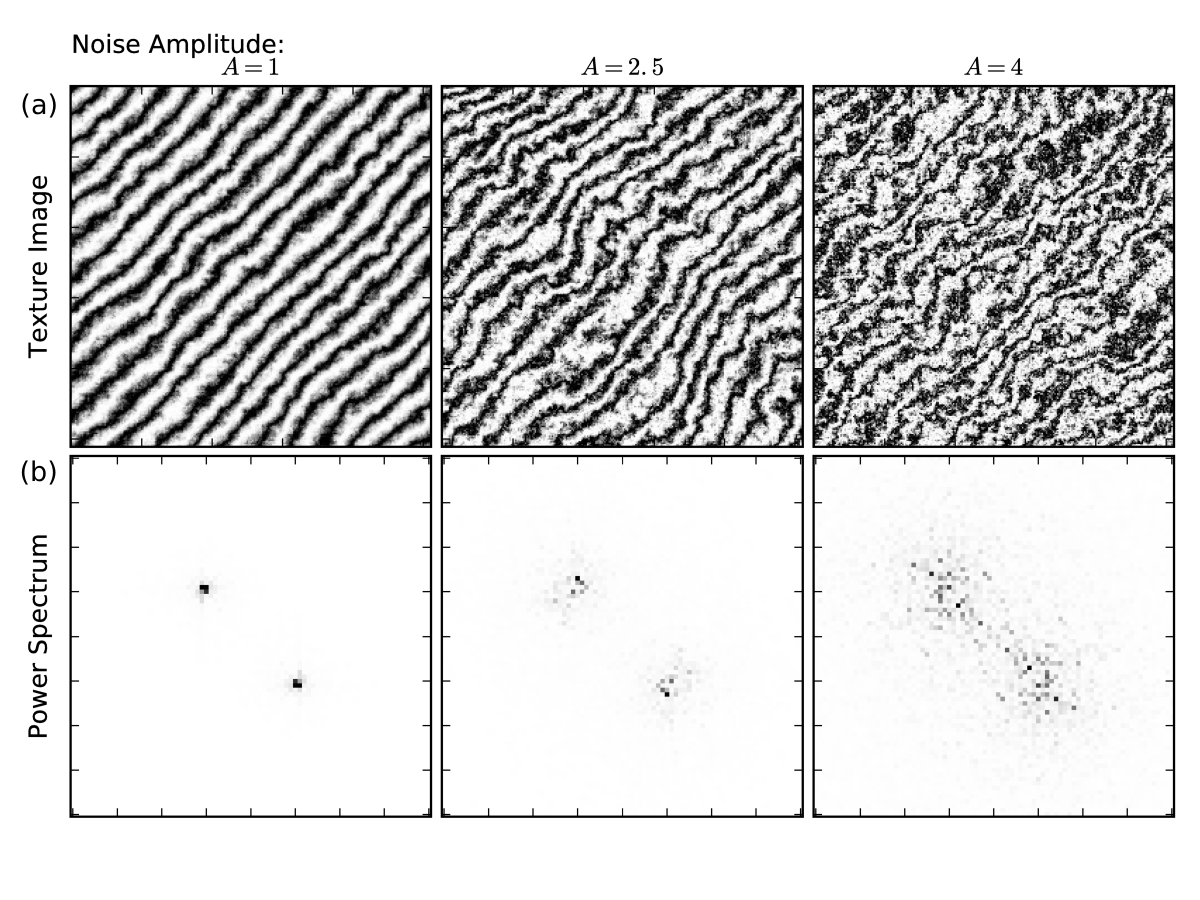}
\caption{Row (a): Synthetic microstructures with scale parameter
  $k=15$ and varying noise amplitudes $A$. Row (b): Associated power spectrum,
  zoomed to relevant region of Fourier space. To aid
  visualization, the intensities are scaled by factors of
  approximately 1, 3, and 100, such that their maxima appear equally dark.}
\label{fig:noisesamples}
\end{figure*}

Figure~\ref{fig:noisesamples} illustrates the multi-scale nature of
the set $\{ x_i \}$ of stochastic texture images. At small noise values, the power spectrum is peaked on two Fourier modes. With increasing noise amplitudes, the peaks of the power spectrum broaden.

\subsection{Angle reconstruction task}

Our first task is to reconstruct a 1-D manifold of images of fixed noise amplitude
and scale parameter, but varying angle. For each trial, the scale parameter was fixed to
$k=15$, corresponding to a modulation wavelength of $1/15$ in units of the linear system size. 
The angles $\theta_i$ take values $(i/N) \pi$ for $i \in \{0, 1, \dots N-1\}$. Note that $0 \leq \theta < \pi$ without loss of generality because $\mathbf{\hat n}_\theta = - \mathbf{\hat n}_{\theta + \pi}$ and thus $\theta$ and $\theta + \pi$ are equivalent for our textures, Eq.~\eqref{eq:textures}. In this subsection we explore datasets with varying dataset sizes $N$ and noise amplitudes $A$.
We compute distances between the images via the CNN (Sec.~\ref{sec:cnntextureprocedure}) and PS (Sec.~\ref{sec:pstextureprocedure}) methods, then use MDS (Sec.~\ref{sec:MDSintro}) to map the images into a $D^\ast=2$ embedding space. 

We quantify reconstruction quality as follows: First, we find the center of mass of all points in the embedded space, and use this as the origin. Second, we calculate angles $\varphi_i$ about the origin, which are unique up to a single additive constant $c$.
Finally, we seek a correspondence between the generating angles $\theta_i$ and the learned values $\varphi_i / 2$. The factor of $1/2$ is necessary because $\theta_i$ ranges from $0$ to $\pi$ whereas $\varphi_i$ ranges from $0$ to $2 \pi$.
We select the constant $c$ to minimize the root-mean-square error,
\begin{equation}
\label{eq:rmse}
\mathrm{RMSE} = \sqrt {\frac{1}{N} \sum_i [\theta_i - (\varphi_i + c)/ 2]^2 }
\end{equation}
Once $c$ is optimized, we use the RMSE to measure the reconstruction quality.

\begin{figure*}
  \centering
   \includegraphics[width=0.8\textwidth]{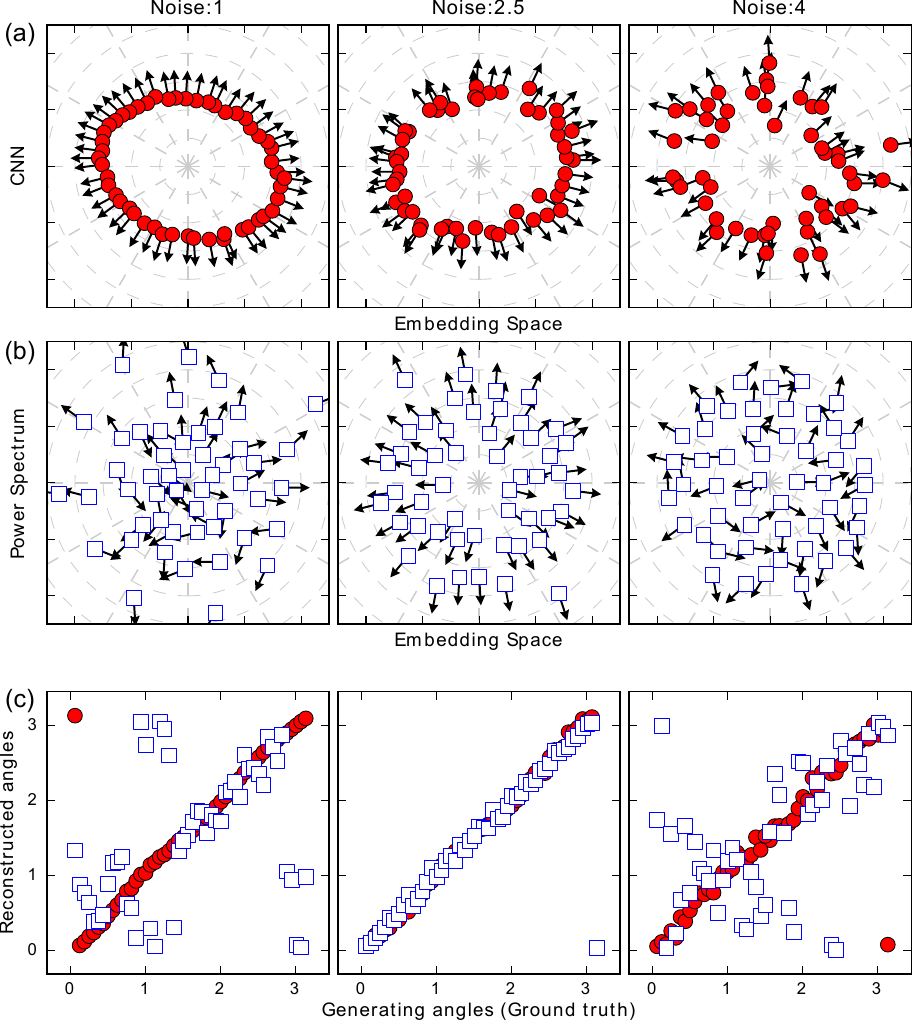}

  \caption{(Color online)  Angle reconstructions. Each dataset contains $N=50$ images with varying angles $\theta_i$, fixed scale $k=15$, and fixed noise amplitude $A \in \{1, 2.5, 4\}$. Red circles correspond to the CNN method (Sec.~\ref{sec:cnntextureprocedure}) and blue squares correspond to the PS method (Sec.~\ref{sec:pstextureprocedure}).
Row (a): MDS embedded points $\hat x_i$
  using the CNN method. Arrows represent $2 \theta_i$ where $\theta_i$
is the angle used to generate image angles. Row (b): MDS
  embedded points $\hat x_i$ using the PS method.
Row (c): Comparison of $\theta_i$ with corresponding angle $(\varphi_i + c)/ 2$ reconstructed from the embedding space. The CNN method yields excellent agreement, and clearly outperforms the PS method.
         }
  \label{fig:fitpanels}
\end{figure*}

Figure~\ref{fig:fitpanels} shows embedded manifolds and corresponding angle reconstructions using dataset size $N=50$ and noise amplitudes $A \in \{1, 2.5, 4\}$.
For low noise amplitudes $A=1$, the CNN distances produce a
ring structure which reflects the generating angles (and associated periodicity) quite
well, whereas the PS method fails. 
For intermediate $A=2.5$, both CNN and PS distances generate good angle reconstructions, but there is much less scatter in the CNN embeddings.
For large $A=4$, the CNN continues to give good angle reconstructions despite scatter in the embedded points, whereas the PS method again fails.
Note that, by construction, the PS method is rotationally symmetric, whereas the CNN method encodes rotational symmetry only approximately. Consequently, the CNN embeddings are somewhat elliptical.

\begin{figure}
  \centering
    \includegraphics[width=0.5\textwidth]{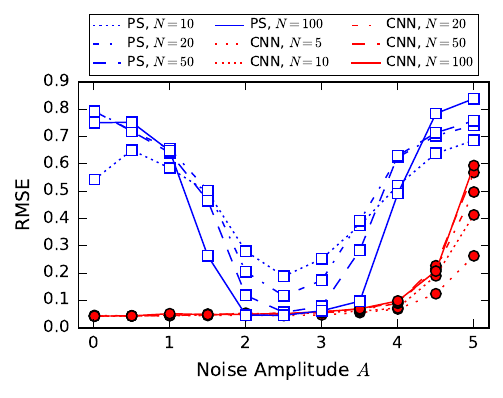}
  \caption{(Color online) The angle reconstruction error, Eq.~\eqref{eq:rmse}, as a function of noise amplitude $A$, with scale $k=15$. Each dataset consists of $N$ images with varying angle $\theta_i$.  The CNN method performs well across a wide range of noise amplitudes $A \lesssim 4.5$, whereas the PS method does best in a narrow range $2 \lesssim A \lesssim 3.5$. Each RMSE estimate represents an average over 100 independent trials.}
  \label{fig:2DRMSE}
\end{figure}

Figure~\ref{fig:2DRMSE} shows the RMSE, Eq.~\eqref{eq:rmse}, for the angle reconstruction task using a variety of dataset sizes $N$ and noise amplitudes $A$.
The CNN embeddings reliably reconstruct the generating angles $\theta_i$ for a wide range of $N$ and $A$. However, the PS
embeddings reconstruct $\theta_i$ only for a
narrow window of $A$, and require a much larger $N$ to reach
comparable accuracy. This behavior can be understood by referring to Fig.~\ref{fig:noisesamples}: At very small $A$, the PS peaks are sharp, and there is little overlap between texture images with different angles. At very large $A$, the PS peaks broaden and exhibit great stochastic fluctuation. The best reconstructions occur at intermediate $A$, for which the peaks have some width but are not dominated by fluctuations, such that PS distances can accurately capture differences in the angle parameter.

\begin{figure}
  \centering
    \includegraphics[width=0.5\textwidth]{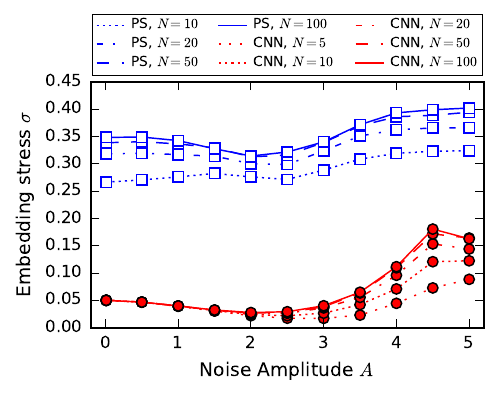}
  \caption{(Color online) The embedding stress $\sigma$ as a function of noise amplitude $A$, with scale $k=15$. Each dataset consists of $N$ images with varying angle $\theta_i$. The CNN method yields much lower stress embeddings than the PS method. }
  \label{fig:2DStress}
\end{figure}

Figure~\ref{fig:2DStress} shows the embedding stress $\sigma$, Eq.~\eqref{eq:MDSstress}, a measure of the fidelity of the MDS embedding.
The CNN embedding exhibits low stress across a wide range of noise amplitudes, whereas the PS distances do not easily embed into a $D^\ast = 2$ embedding space.
The stress of both CNN and PS embeddings grows with the noise amplitude.
We interpret this as follows:
At zero noise, the space of texture images has a single parameter, the
angle. With finite noise, this 1-D manifold of texture images expands
into a much higher-dimensional space. The effective expansion volume
increases monotonically with the noise amplitude. Consequently, it
becomes increasingly difficult to embed this very high-dimensional
manifold using $D^\ast = 2$, which is reflected in the increasing
embedding stress. 

\subsection{Three dimensional manifold reconstruction task}
\label{sec:threedeetask}

Here we embed texture images from all three generating parameters shown in Fig.~\ref{fig:3DManifoldScheme}:
the angle $0 \leq \theta < \pi$, scale $5 \leq k < 15$, and noise amplitude $0.5 \leq A < 2$.
We generate a dataset of $N=1000$ texture images by varying each parameter through $10$ equally spaced increments.
As before, we determine the distances
between images using CNN (Sec.~\ref{sec:cnntextureprocedure}) and power spectrum (Sec.~\ref{sec:pstextureprocedure}) methods,
then use MDS (Sec.~\ref{sec:MDSintro}) to embed these distances into spaces of varying dimension $D^\ast$.

\begin{figure}
  \centering
\includegraphics[width=0.5 \textwidth]{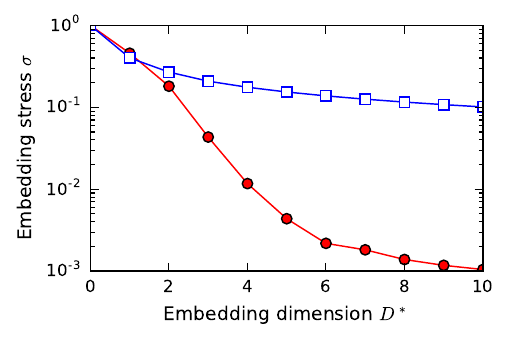}
\caption{(Color online) The stress $\sigma$ as a function of embedding dimension $D^\ast$ for a dataset of $N=1000$ images with varying $A$, $k$, and $\theta$.
Stress from the CNN method (red) decreases approximately exponentially for $D^\ast \lesssim 6$.
Stress from the PS method (blue) decreases much more slowly. That is, the PS distances do not naturally embed into low dimensions.
}
  \label{fig:3DStress}
\end{figure}

Figure~\ref{fig:3DStress} shows the embedding stress $\sigma$ as a
function of embedding dimension $D^\ast$. We observe
a much lower stress using the CNN distances. The stress $\sigma$ decays exponentially up to about $D^\ast = 6$ and flattens soon after.
That is, with $\approx 6$ descriptors per image, MDS has learned a representation of the texture images quite faithful to the CNN distances.
Conversely, for the PS method, $\sigma$ decays very slowly with $D^\ast$,
suggesting that there is no natural low-dimensional embedding manifold.

\begin{figure*}
    \centering
\includegraphics[width=0.8 \textwidth]{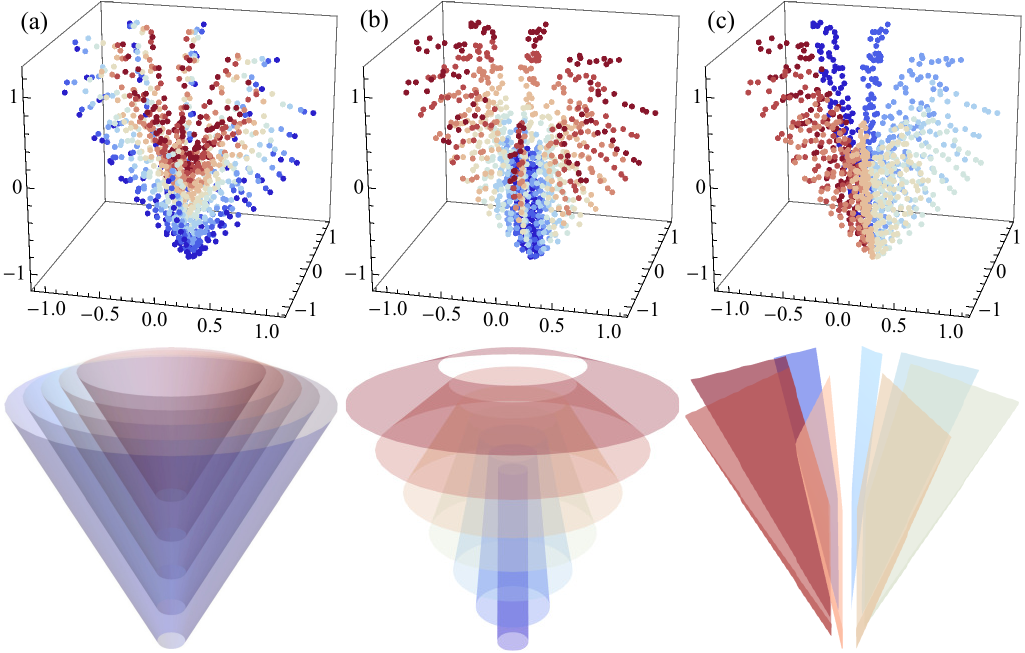}

    
  \caption{(Color online) The $D^\ast = 3$ embedding of a dataset with
    $N=1000$ image  of varying noise amplitude $A$,
  scale $k$, and angle $\theta$. Columns (a), (b), and (c) are colored
  by $A$, $k$, and $\theta$, respectively. For each column, the top panel shows
    the embedded points, and the bottom panel displays a schematic representation of surfaces with constant generating parameter.}
  \label{fig:3DVis}
\end{figure*}

\begin{figure}
  \centering
    \includegraphics[width=.4\textwidth]{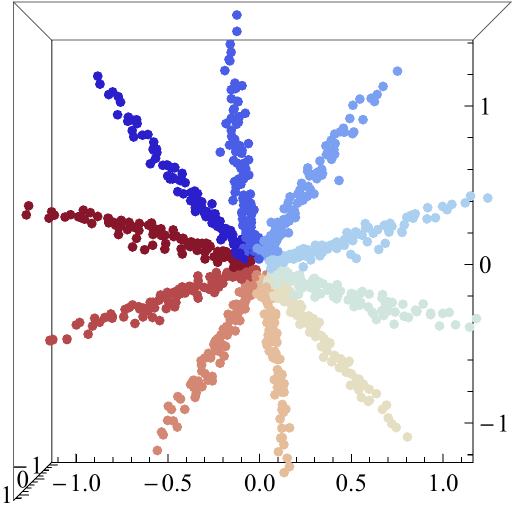}
  \caption{(Color online) Axis-aligned view of the $D^\ast = 3$ embedding colored by
    angle $\theta$ (cf. Fig.~\ref{fig:3DVis}), demonstrating strong correspondence with angles in the embedding space.}
  \label{fig:3DAngles}
\end{figure}

The panels in Fig.~\ref{fig:3DVis} show the CNN method embeddings in a $D^\ast = 3$ space.
The generating parameters $A$, $k$, and $\theta$ emerge as a nonlinear coordinate system that spans a roughly conical solid in the embedding space.
Column (a) of Figure~\ref{fig:3DVis} shows surfaces in which the noise amplitude $A$ is held fixed, and the parameters $\theta$ and $k$ are allowed to vary. For each $A$, the embedded points form an approximately conic surface. Cones with larger $A$ are nested inside of cones with smaller $A$.
Column (b) of Figure~\ref{fig:3DVis} shows surfaces in which the scale $k$ is held fixed, allowing $A$ and $\theta$ to vary. Surfaces of constant $k$ are harder to describe, but resemble fragments of conic surfaces with varying angles. Surfaces with smaller $k$ are nested inside of ones with larger $k$.
Column (c) of Figure~\ref{fig:3DVis} shows surfaces in which the angle $\theta$ is held fixed, allowing $A$ and $k$ to vary. Rotating to a top-down view (Fig.~\ref{fig:3DAngles}), one observes that $\theta$ is very well captured by the azimuthal angle of the cone structure.

\begin{figure*}
    \centering
\begin{subfigure}[b]{0.3\textwidth}
        \includegraphics[width=\textwidth]{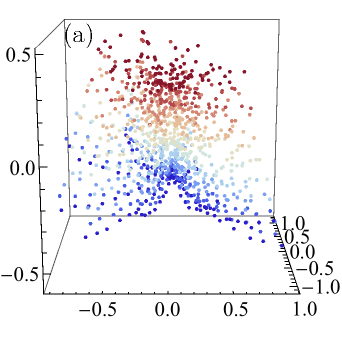}
    \end{subfigure}
\begin{subfigure}[b]{0.3\textwidth}
        \includegraphics[width=\textwidth]{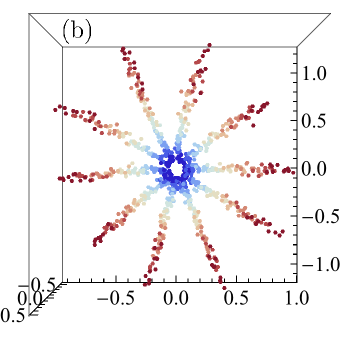}
    \end{subfigure}
\begin{subfigure}[b]{0.3\textwidth}
        \includegraphics[width=\textwidth]{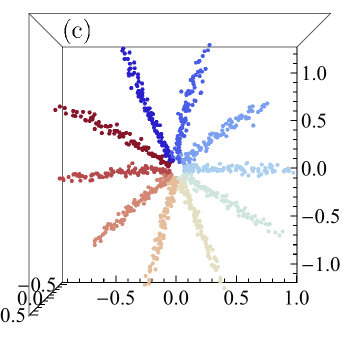}
    \end{subfigure}
\caption{(Color online) The $D^\ast = 4$ embedding of a dataset with
  $N=1000$ images of varying noise amplitude $A$,
  scale $k$, and angle $\theta$. Panels (a), (b), and (c) are colored
  by $A$, $k$, and $\theta$, respectively.
To visualize the data, we select a 3-D projection that illustrates a decoupling of the generating parameters, in which they manifest as roughly cylindrical coordinates:  $A$ maps to longitudinal height, $k$ to radius, and $\theta$ to azimuthal angle.}
\label{fig:4DEmbedding}
\end{figure*}

Figure~\ref{fig:4DEmbedding} shows a 3-D projection of the $D^\ast = 4$ embedding.
In this projection, we observe that the generating parameters appear approximately as
cylindrical coordinates. The noise parameter $A$ appears approximately
linearly as a longitudinal coordinate.

To quantitatively evaluate the quality of embeddings for general
dimension $D^\ast$, we performed linear regression to model
the parameters $A_i$ and $k_i$ for each embedding point $\hat{x}_i$:
\begin{eqnarray}
\hat{A}_i = \beta_A \cdot \hat{x}_i + \gamma_A \, ,\\
\hat{k}_i = \beta_k\cdot\hat{x}_i + \gamma_k \, .
\end{eqnarray}
The regression vectors $\beta_p$ and scalars $\gamma_p$ for $p \in \{A, k\}$ were found using ordinary least
squares by minimizing $\sum_i (p_i - \hat p_i)^2$. We then assess fit quality using the coefficient of
determination,
\begin{equation}
R^2 = 1 - \frac{\sum_i ( p_i - \hat{p}_i )^2}{\sum_i (p_i-\bar{p} )^2},
\end{equation}
where  $\bar{p} = \sum_i p_i / N$.
An $R^2$ value of $1$ indicates a perfectly linear
relationship. Table~\ref{tab:regression} shows $R^2$ values for
$k$ and $A$ in various embedding dimensions. In particular, $R^2 \approx 1$
is achieved already with $D^\ast = 4$ and higher-dimensional embeddings ($D^\ast > 4$)
 yield only marginal increase in fit quality.

\begin{table}
\centering
\begin{tabular}{ c|cccccccccc }
  Embedding dimension $D^\ast$ & 2 & 3 & 4 & 5 & 6 & 10 & 50\\\hline
 $R^2$ for scale $k$  & .532 & .720 & .901 & .916 &.916 & .930 &
                                                                 .980\\
$R^2$ for noise amplitude $A$ & .183 & .231 & .908 & .950 & .951 &
                                                                   .972 & .983
\end{tabular}
\caption{
The coefficients of determination $R^2$ for linear regression models that map embedded points $\hat x_i$ to $k_i$ or $A_i$.
For embedding dimensions $D^\ast \geq 4$, the linear models achieve $R^2$ values near the ideal of 1.
}
\label{tab:regression}
\end{table}

\begin{figure}
    \centering
\begin{subfigure}[b]{0.4\textwidth}
        \includegraphics[width=\textwidth]{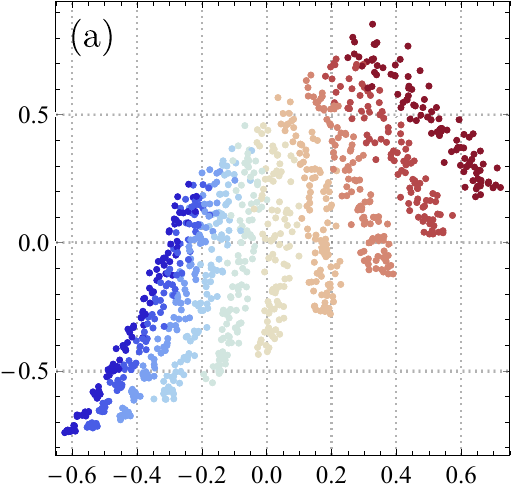}
    \end{subfigure}
\begin{subfigure}[b]{0.4\textwidth}
        \includegraphics[width=\textwidth]{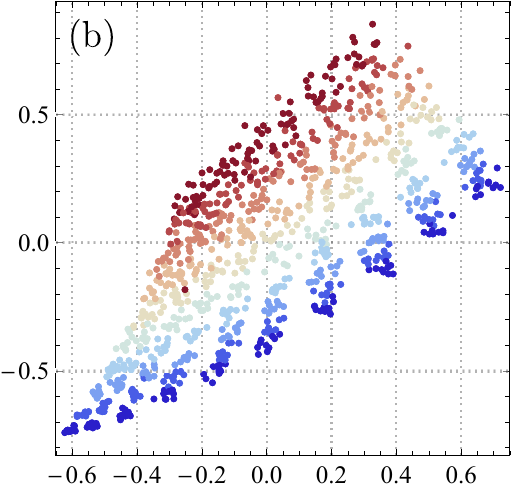}
    \end{subfigure}
\caption{(Color online) The 2-D projection, colored by (a) scale and (b) noise amplitude, of the $D^\ast = 6$
  embedding of a dataset with $N=1000$ images of varying $A$, $k$, and
  $\theta$. This projection was selected using linear regression
  to find the two dimensions that best capture the scale $k$ and noise
  amplitude $A$ parameters.}
  \label{fig:6dregression}
\end{figure}

Figure~\ref{fig:6dregression} shows the $D^\ast = 6$
embedding projected onto the two dimensions,
$\{\beta_k,\beta_A\}$, that
best linearly model the noise amplitude and scale parameters. The scale
and noise parameters are nearly orthogonal; the angle between
$\beta_k$ and $\beta_A$ is
\begin{equation}
\theta = \arccos \left(\frac{\beta_A \cdot
  \beta_k }{ \|\beta_A\|\|\beta_k\|}\right) = 91.0 \degree.
\end{equation}

\section{Discussion}

The effectiveness of the PS on the angle reconstruction task (see Fig.~\ref{fig:2DRMSE}) is best for moderate values of the
noise, but weak outside of this window. Although we discussed the
mechanisms at play, this result is at first counterintuitive; The images consist of perturbed sinusoidal stripes that
coincide well the Fourier basis used by the PS. The performance of the CNN texture vector is much better, achieving
high accuracy and consistent performance across the spectrum of
noise. This can be attributed to several advantages of the CNN-based
approach.

First, the CNN uses local filters as opposed to global modes. Global
features can suffer from interference effects, where similar small
scale features which appear at a large distances from each other can
add together destructively. Local features do not suffer from this
type of failure mode, and so are more robust to noisy variations in
patterns. This is similar to advantages of compact support in wavelet
approaches to signal processing, which are well studied~\cite{Mallat08}.

Second, compositions of  convolutional filters and nonlinear
activations represent very non-trivial correlations between the pixels
within their receptive fields, so that individual neurons are
sensitive to higher order statistics that are not captured
in the PS representation. For example, higher order statistics can
directly characterize complex features such as domain edge curvature. 

Lastly, the pooling layers in the CNN operate similarly to
coarse-graining in physics, and is designed to capture relevant system characteristics while discarding
unimportant ones~\cite{[{ An explicit correspondence between the
    variational renormalization group and deep learning using
    Boltzmann Machines is demonstrated in }][{}]Mehta14}.
In the CNN, repeated convolutional and pooling operations effectively implement coarse-graining over multiple
layers of abstraction.
Features that appear in deeper layers (\emph{i.e.} further from the input) of the CNN have a  spatially larger receptive field, and are more robust to small changes of the input due to the coarse-graining.
Thus, larger-scale CNN features are naturally insensitive to
smaller-scale texture details, which we believe is key to microstructure
analysis as well as computer vision tasks.
A trade-off with deep neural networks is that it can be difficult to understand concretely what a particular
activation in a CNN represents; however, this is an active area of
research~\cite{Erhan10,Yosinski15,Selvaraju16}.

\section{Conclusions and future directions}

We have introduced a method for unsupervised detection of the
low-dimensional structure of a distribution of texture images using
CNNs. We discuss the uses of this as a framework for the analysis of materials
microstructure to learn dimensionality and topology of microstructure
families using low-dimensional quantitative descriptions of
microstructure. Compact microstructure characterization forms a platform
for the construction of reduced order models that connect processing
to microstructure, and microstructure to properties. This approach is
applicable to small data sets, which is an important design
factor in materials science and other disciplines where acquiring data
can be expensive. In this work, we apply manifold learning to a synthetic
dataset. This controlled context enables us to quantify the success of manifold learning. We anticipate that similar manifold learning approaches will prove effective for follow-on studies of real materials. For example, DeCost \emph{et al.} recently demonstrated success in mapping the microstructures of ultra high carbon steels~\cite{DeCost17}.

The method presented in this work is computationally efficient. In our Theano implementation, running on a single GPU, it takes about a millisecond to compute the distance between texture vectors that represent two images. MDS operates on all distance pairs, and thus scales quadratically with the size of the dataset. The MDS calculation on our full dataset of $10^3$ synthetic images (Sec.~\ref{sec:threedeetask}) completed in about 30 minutes. The dominant cost was the MDS embedding procedure, which took about 18 minutes. Calculating the $\approx 10^6/2$ texture vector distances took about 12 minutes. 

A limitation of the transfer learning approach is that it requires a
well-trained CNN with applicability to the target domain, which
presently limits our analysis to 2-D micrographs. One path for
improvement is to directly train CNNs on a large database of standardized microstructure images.
Such a database could also be used to develop latent variable models
(e.g.~\cite{Kingma13}) that would reflect the microstructural
generation process. These end-to-end models would
enable direct inference of low-dimensional generating parameters
and direct generation of new microstructure image samples.

The work of Ref.~\onlinecite{Ustyuzhaninov16} suggests that, instead of using transfer learning on natural images, it may be possible to characterize microstructure textures using randomized CNNs. Specifically, Ustyuzhaninov \emph{et al.} conclude that suitably structured random, shallow, multiscale networks can, in some cases,
be used to generate higher quality textures than those generated from
a trained CNN. However, Fig.~1 of Ref.~\onlinecite{Ustyuzhaninov16} shows that
 the distance matrix generated from the trained CNN is closer to the identity compared to the distance matrix generated from the random network. This suggests that the trained CNN is a better starting point for comparing texture images.
The capability of random
networks to perform low-dimensional embeddings of microstructures remains an open question. The use of random networks suggests exciting opportunities to operate on other data
modalities, \emph{e.g.} three-dimensional microstructure data~\cite{Rollett07,Sundararaghavan05,Xu14} and/or grain orientation data~\cite{Adams93,Humphreys01,Orme16}, both of which are outside the domain of natural image
characterization.

Lastly, we consider the rotation group, which factors into
microstructure analysis in at least two ways. Firstly, rotations appear through spatial
transformations of the image plane. The standard CNN architecture
does not explicitly incorporate such transformations. This is evident
in the small but consistent biases in the angular reconstruction task
(see Figs.~\ref{fig:fitpanels} and~\ref{fig:2DRMSE}).  However, the relatively strong
performance of the network on this task indicates that the network has
implicitly learned approximate representations of the rotation group
during the training procedure. A second way that the rotation group appears is in grain orientation data. CNNs designed to process RGB images do not directly represent the group structure of crystalline orientations. Further texture
characterization work might explicitly incorporate the action of
rotations, \emph{e.g.} building upon
Refs.~\onlinecite{Mallat12,Cohen14,Gens14,Dieleman16,Cohen16}.

\begin{acknowledgments}
We acknowledge funding support from a Laboratory Directed Research and Development (LDRD) DR (\#20140013DR), and the Center for Nonlinear Studies (CNLS) at the Los Alamos National Laboratory (LANL).
We also thank Prasanna Balachandran and James Theiler
for useful discussions and feedback.
\end{acknowledgments}

\appendix
\section{Attribution of microstructure images}
\label{sec:imagecredits}

The images appearing in Fig.~\ref{fig:texturereconstructions} are released in the public domain, and available online. Permanent links are as follows. Top row, from left to right:
\begin{enumerate}
\item \url{https://commons.wikimedia.org/w/index.php?title=File:Gailbach-Tonalit.jpg&oldid=179586143}.
\item \url{https://commons.wikimedia.org/w/index.php?title=File:PearliteSph3.jpg&oldid=149930090}.
\item \url{https://commons.wikimedia.org/w/index.php?title=File:Martensit.jpg&oldid=144193548}.
\item \url{https://commons.wikimedia.org/w/index.php?title=File:Lamine316L.jpg&oldid=66965939}.
\end{enumerate}

Bottom row, from left to right:
\begin{enumerate}
\setcounter{enumi}{4}
\item \url{https://commons.wikimedia.org/w/index.php?title=File:Microstructure_of_rolled_and_annealed_brass;_magnification_400X.jpg&oldid=144992203}.
\item \url{https://commons.wikimedia.org/w/index.php?title=File:Ferrite-perlite-steel-A285.jpeg&oldid=140529933}.
\item \url{https://commons.wikimedia.org/w/index.php?title=File:Pearlite1.jpg&oldid=184321025}.
\item \url{https://commons.wikimedia.org/w/index.php?title=File:Feuerverzinkte_Oberfl\%C3\%A4che.jpg&oldid=140580291}
\end{enumerate}

\bibliography{../../PRE-proof/gmetricbib.bib}

\end{document}